\documentclass[10pt,a4paper,aps,twocolumn,showpacs,superscriptaddress,nofootinbib]{revtex4-1}
\usepackage[colorlinks,hyperindex]{hyperref}
\usepackage{epsfig}
\usepackage{float}
\usepackage{float,upgreek,yfonts}
\RequirePackage{amssymb,amsmath}
\usepackage[latin1]{inputenc}
\usepackage{graphicx}
\usepackage{indentfirst}
\usepackage{color}
\begin{document}
\title{Baryon production probability via the nuclear configurational entropy}
\author{G. Karapetyan}
\email{gayane.karapetyan@ufabc.edu.br}
\affiliation{Centro de Ci\^encias Naturais e Humanas, Universidade Federal do ABC - UFABC, 09210-580, Santo Andr\'e, Brazil}
\begin{abstract}
We study the net-baryon production at forward rapidities within the
Color Glass Condensate paradigm. At high energy regime, the leading baryon production mechanism is shown to change from recombination to independent fragmentation.
The nuclear configurational entropy (NCE) constructed upon forward scattering amplitudes,  allows to predict the two free parameters
that govern the anomalous dimension of the target gluon distribution. The global minimum of the NCE indicates a point of  stability in pp/pA/AA collisions at LHC energies, corroborating and matching HERA data for hadron spectra measured in pp and dAu collisions at RHIC energies, with accuracy between 1.2\% and 1.8\%, respectively for the two free parameters.

\end{abstract}
\maketitle
\section{Introduction}

Recently, within the theory of the strong interactions, the so-called Color Glass Condensate (CGC) and QCD, the configurational entropy (CE) approach is widely used to describe the vast bulk of new experimental data, as well as the set of the parameters during the theoretical calculations \cite{Karapetyan:2018yhm}.
A lot of information has been accumulated by using CE approach concerning nuclear reactions under extremal conditions as a high excited state \cite{Karapetyan:plb, Karapetyan:2020epl}.
Using such a technique, a huge bulk of information, that has been  obtained from a wide range of experimental data and numerical theoretical calculation, already proved the sustainability of the CE approach as a robust method applied to the strong interactions, in order to find the critical points of the CE underlying hot nuclear systems.  Such critical points of the CE, summing up stability and dominance of the system states that correspond to these critical points of the CE, give us a hint about the mechanism of the interaction and the further progress of  various excited states of the system.
Recently, the CE principle was widely used to study such important aspects of the collisions as the production of scalar and tensor mesons \cite{Ferreira:2019inu,Bernardini:2018uuy,Braga:2018fyc,Bernardini:2016hvx,Barbosa-Cendejas:2018mng,Ferreira:2019nkz}, glueballs \cite{Bernardini:2016qit}, charmonium and bottomonium  \cite{Braga:2017fsb}, the quark-gluon plasma \cite{daSilva:2017jay}, barions \cite{Colangelo:2018mrt} as well as various questions within
 AdS/QCD \cite{Ma:2018wtw}.
The CE approach, which has been established in a number of the studies \cite{Gleiser:2012tu, Gleiser:2011di, Gleiser:2013mga} and which was a logical continuation of Shannon's information entropy, has been successfully applied in  Refs.
\cite{Gleiser:2018kbq,Sowinski:2015cfa,Gleiser:2014ipa,Gleiser:2015rwa}.
This approach also has been properly employed to find the critical points of the various physical systems in Refs. \cite{Casadio:2016aum,Fernandes-Silva:2019fez,Braga:2016wzx, Bernardini:2019stn,Braga:2020myi,Alves:2017ljt,Alves:2014ksa}, for AdS black holes and their quantum portrait as Bose--Einstein graviton condensates, emulating magnetic structures \cite{Bazeia:2018uyg}.
Using the CGC paradigm as a convenient framework in order to check the CE sustainability one can expect to obtain successful results in the range of high energy particle physics as it was presented in a number of studies where the parton saturation effect has been suggested \cite{Karapetyan:2018yhm,Karapetyan:2019ran,Karapetyan:2019epl,Ma:2018wtw,Karapetyan:plb,Karapetyan:2017edu,Karapetyan:2016fai,Lee:2017ero}.
In the above-mentioned studies, the cross section of the hadron interaction has been used during the calculation of the CE functions, as a spatially localized configuration feature of the hot interaction system.
It allowed to calculate the Fourier transform of the reaction cross section  in order to obtain the nuclear modal fraction and, subsequently, the critical points of the CE \cite{Karapetyan:2018yhm,Karapetyan:plb,Karapetyan:2017edu,Karapetyan:2016fai}.
Such critical points, within the QCD approach, can be used as a starting point for explaining and understanding the whole spectrum of observed phenomena, as well as to derive the right parameters for theoretical calculations.

The area of strong interactions including the LHC  is the outstanding possibility to check and prove a variety of striking observed phenomena.
The whole spectrum of such phenomena is explained within the high-density parton CGC approach at LHC energy range.
In the low $x$ regime,
which corresponds, at lowest order in perturbation theory,
to the longitudinal momentum fraction carried by a parton in the hadron,
the QCD evolution of the CGC describes the RHIC and LHC data for hadron production.
The probability of baryon production can be expected in both the central and the forward rapidity regions.
It is suggested that the forward baryons arise from the recombination of the three valence quarks with the target mainly from gluons and sea quarks. In other words, in the high energy regime at large rapidities, baryons are produced
during the interaction of relatively large fractional momentum partons of the projectile ($x_{1}$) with low fractional momentum partons in the target ($x_{2}$) \cite{Duraes}.
At the given energy regime one can consider the target as a very dense system of partons, which consist mostly of gluons.
Therefore, it may be observed as the CGC, characterized by a state with a very high density of partons. In such a state the only effects that can cause the global redistribution of the partons are the nonlinear effects of QCD. Thus, one can expect the change in the production cross section and be interpreted from the point of view the entropy conservation.
In deeply inelastic scattering (DIS) the CGC formalism is based on
a so-called saturation scale, $Q_s$. It is a specific momentum scale that grows with the reaction energy and as $x$ is decreasing.
This state is saturated for gluons with transverse momenta and defines the onset of nonlinear (or saturation) effects.
According to the CGC formalism, at the certain projectile energy,
the valence quarks obtain huge transverse momentum, lost its coherence, thus the mechanism of the leading baryon production changes from recombination to independent fragmentation.
The present paper uses the CE as a convenient and complete tool in order to justify the leading baryon production in pp/pA/AA collisions at LHC energies within the CGC formalism.

This paper is organized as follows.
We give a brief description of the CGC formalism together with the basic formulas and parameters in the Sect. \ref{s2}.
It includes also the forward dipole scattering amplitude model of the calculations.
Sect. \ref{s3} is devoted to results of the calculations, via the nuclear CE approach, and give our predictions for the probability for baryon production in proton-proton or/and proton-nucleus collisions, analyzing the parameters that govern the anomalous dimension of the target gluon distribution, composing the forward scattering amplitude that constructs the nuclear cross section. 
These parameters will be derived from the critical point of the nuclear CE.
Finally, in the Conclusion, we summarize our main results.

\section{Baryon production probability in the framework of CGC}
\label{s2}
The main characteristics of $pp/pA/AA$ collisions at LHC energies within
the CGC formalism is the probability for the leading baryon production,
which can be represented by the differential cross section of a hadron with a
transverse momentum $p_T$ and a rapidity $y$ in the following form \cite{Duraes,buw}:
\begin{equation}
\label{PE}
\frac{dN}{d^2p_Tdy}= \frac{1}{(2\pi)^2 } \int_{x_{\text{F}}}^1 \frac{dz}{z^2} \frac{D(z)}{q_T^2}\;x_1q_v(x_1)\;\varphi\left(x_2, q_T \right).
\end{equation}

In Eq. \eqref{PE}, $D(z)$ is the net-baryon fragmentation function, which can be expressed by the formula:
\begin{equation}
D(z) \equiv D_{\Delta B/q}(z)=D_{B/q}(z)-D_{\bar B/q}(z),
\end{equation}
where the symbol $B$ is referred to the produced baryon, and $z = E_{B}/E_{q}$ is the fraction of the baryon energy taken from the fragmenting quark energy ($E_q$); $x_1=q_{T} \, e^{y}/\sqrt{s}$ and
$x_2=q_{T} \, e^{-y}/\sqrt{s}$ are the fractional momenta of the projectile quark and the target gluon, respectively.
One also defines the quark transverse momentum in Eq. \eqref{PE} as $q_T=\sqrt{p_T^2+m^2}/z$, and the Feynman variable as
$x_F = \sqrt{p_{T}^2+m^2} \, e^{y}/\sqrt{s}$.
The product  $x_1 \, q_v(x_1)$ represents the valence quark distribution of the projectile hadron.
The unintegrated gluon distribution of the hadron target can be expressed by the following expression, 
\begin{equation}
\varphi(x_{2},q_{T}) = 2\pi q_{T}^{2} \int {r_{T}dr_{T} \mathcal{N}(x_{2},r_{T}) J_{0}(r_{T}q_{T})}
\end{equation}
where  $J_0$ is the zeroth order Bessel function and  $\mathcal{N}(x_{2},r_{T})$ denotes the forward scattering amplitude; $r_T$ is a radius of a
color dipole off a hadron target. As digressed in \cite{buw}, one can describe the HERA data for hadron spectra measured in pp and dAu collisions at RHIC energies. 
The forward scattering amplitude contains the information about the non-linear quantum effects in the hadron wave function and can be
represented by a simple Glauber-like formula as \cite{buw}
\begin{equation}
{\cal{N}}(x,r_T) = 1 - \exp\left[ -\frac{1}{4} (r_T^2 Q_s^2)^{\gamma (x,r_T^2)} \right] \,,
\label{ngeral}
\end{equation}
where $\gamma(x,r_T)$ regards the anomalous dimension of the target gluon distribution,  reading 
\begin{equation}
\gamma(x,r_T)=0.628+0.372\frac{(\omega^a-1)}{\omega^a-1+b}.\label{ghj}
\end{equation}
Here $\omega  \equiv (r_TQ_s(x))^{-1}$.  

Hence, the scattering amplitude (\ref{ngeral}) will be used to construct the cross section of this system,
whose nuclear CE will be then computed, as a function of the parameters $a$ and $b$ in Eq. \eqref{ghj}. 

According to Ref. \cite{Kharzeev}, at interaction fixed energy, $\mathcal{E}$,
and the rapidity value, $y$, the saturation scale can be represented by
\begin{equation}
\label{QSWY}
Q^2_s(A,y,\mathcal{E}) = Q_0^2(A,\mathcal{E}_0) \left( \frac{\mathcal{E}}{\mathcal{E}_0} \right)^{{\lambda}} e^{{\lambda} y}
\end{equation}
with
\begin{equation}
{\lambda} \equiv \frac{d\log(Q_s^2(x)/\Lambda^2_{\rm})}{d\log (1/x)}\approx  0.252.\label{f1f}
\end{equation}
This value was obtained in Ref. \cite{Karapetyan:2019ran}, also in the context of the CE, in hadron-nucleus collisions. Hadron multiplicities and $Au$ nuclei were studied to derive the CE as a function of the saturation scale ruling deep inelastic scattering phenomena in QCD. This obtained value of $\lambda$ match experimental data within $\sim 0.4\%$.
One can obtain a final expression, by integrating
\eqref{QSWY} over all possible values of $y$:
\begin{equation}
\label{QSR}
Q^2_s(\mathcal{E}) = \Lambda^2_{\rm }e^{\sqrt{2 \delta \log\left(\frac{\mathcal{E}}{\mathcal{E}_0}\right) + \log^2\left[Q^2_s\left(\frac{\mathcal{E}_0}{\Lambda^2_{\rm }}\right)\right]}},
\end{equation}
where  $Q^2_s(\mathcal{E}_0)$ represents the saturation scale at  energy $\mathcal{E}_0$
with the value of the parameter $\Lambda^2_{\rm } = 0.04 {\rm GeV}^2$
and  $\delta = \lambda   \log(Q^2_{s0}/\Lambda^2_{\rm })$, where $Q^2_{s0}$ is the saturation scale at $y = 0$.

Two fitted free parameters $a=2.82$ and $b=168$ describe satisfactory the RHIC data on hadron production \cite{Duraes}.
It should be noted, however, that the model satisfactorily describes the nonlinear approach, which corresponds to the small $q_T$ and does not include any expectation for the large value of the transverse momentum of the hadron. The aim of the next section is to obtain
the CE, underlying the forward scattering amplitude \eqref{ngeral},

\section{Leading baryon production and configurational entropy}
\label{s3}
Our focus in this paper is on high energy net-baryon rapidity distribution of $pp/pA/AA$ collisions at LHC energies in QCD via the CE paradigm as the most appropriate technique in order to get the whole picture of the interaction.
As it was shown in Ref. \cite{Bernardini:2016hvx}, it is possible to decompose the Fourier function associated with the baryon production probability into a series of weighted nuclear modes.
This procedure brings to the specification of the normalized structure factor via collective coordinates.
In order to increase the information package about the hadron interaction at high energy range, one can use the modal fraction of the CE, and connect the probability distribution with the power spectrum of the baryon rapidity distribution.
Any localized system has a set of critical points that reflect all information about the mechanism of nuclear interaction and can be considered via the Shannon's information entropy.
As a result, one can expect to obtain an adequate explanation of the observed phenomena from CE and prepare the correct and useful parameters and variables, which describes the interaction systems.

The nuclear CE is based on the assumption of the localized functions \cite{Gleiser:2012tu,Gleiser:2013mga}.
With Eq. \eqref{ngeral} describing the forward scattering amplitude, the cross section, $\sigma_\mathcal{N}({x,a,b})$ is defined as
\begin{equation}
\sigma_\mathcal{N}({x,a,b})=|{\cal{N}}(x,r_T)|^2,
\end{equation}
taking into account the the anomalous dimension of the target gluon distribution, Eq. (\ref{ghj}).
The probability of baryon production is transformed via the energy-weighted correlation Fourier function, using Eq. (\ref{PE}), which is  a localized function:
\begin{equation}
\label{34}
\sigma_\mathcal{N}({k,a,b})=\frac{1}{2\pi}\! \int_{\mathbb{R}}\sigma_\mathcal{N} (x,a,b)\, e^{ikx} d x.
\end{equation}  In such a way, one can represent the nuclear modal fraction as the following:
\begin{equation}\label{modall}
f_{\sigma_\mathcal{N}({k,a,b})}=\frac{\vert \sigma_\mathcal{N}({k,a,b}) \vert^2}{\int_{\mathbb{R}}\vert \sigma_\mathcal{N}({k,a,b})\vert ^2 dk}.
\end{equation}
After this, the next step is to apply appropriate expression for the CE \cite{Gleiser:2012tu} in order to calculate the corresponding critical points:
\begin{equation}
\label{333}
{\rm CE}(a,b) =  - \int_{\mathbb{R}} f_{\sigma_\mathcal{N}({k,a,b})} \log  f_{\sigma_\mathcal{N}({k,a,b})}d k.
\end{equation}
Thus, having as a starting point the net-baryon rapidity distribution of $pp/pA/AA$ collisions \cite{Duraes} given by Eq. (\ref{PE}), we can  compute the nuclear CE, using Eqs. (\ref{34}) -- (\ref{333}).

\begin{figure}[!htb]
       \centering
                \includegraphics[width=2.9in]{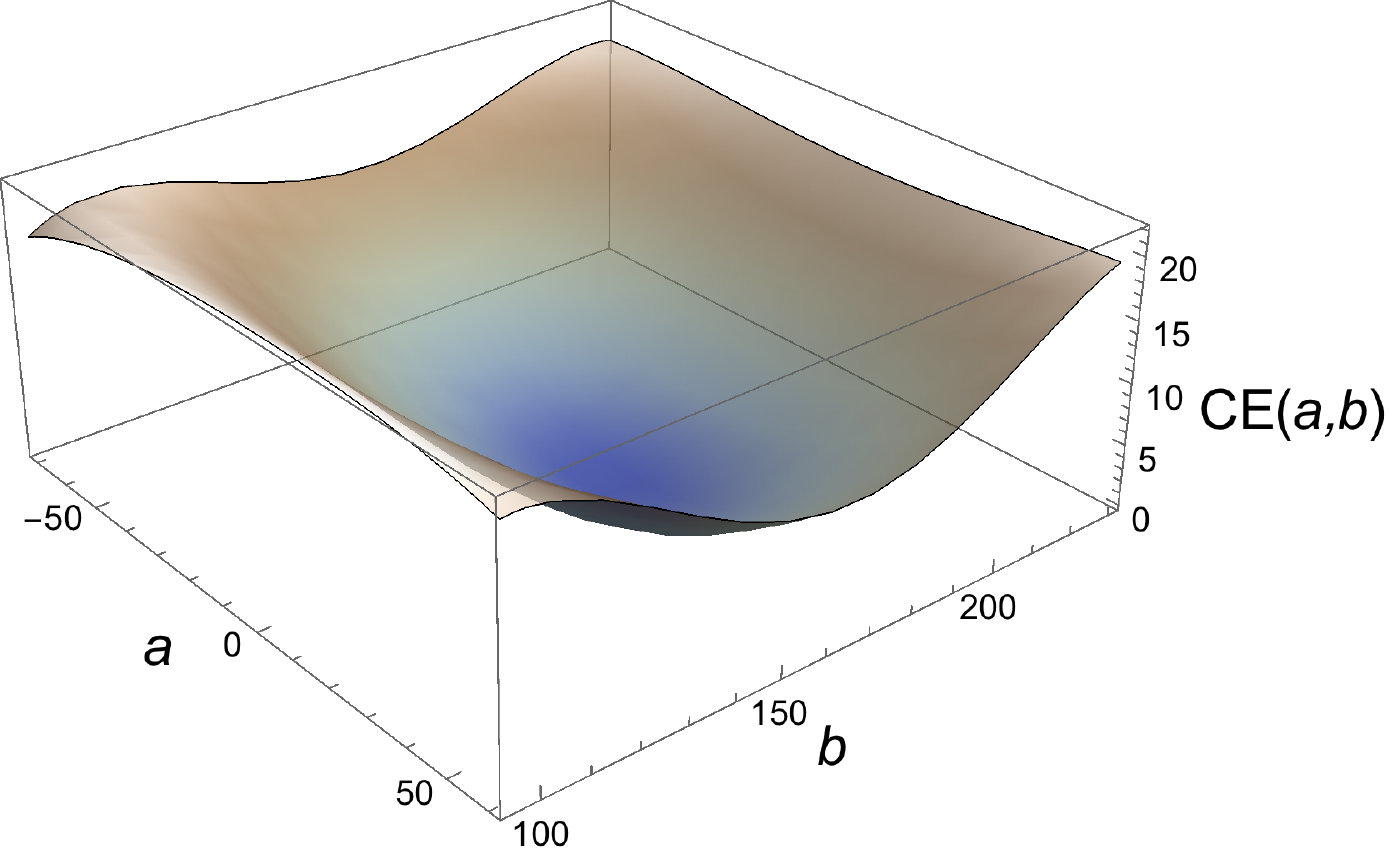}
                \caption{Nuclear CE as a function of the free parameters $a$ and $b$, for the global minimum CE($a,b$) = 0.476, for $a=2.77$ and $b=166.02$. These values matching HERA data with accuracy of 1.8\% for the parameter $a$ and 1.2\% for the parameter $b$, correspondingly.}
                \label{fff1}
\end{figure}

\begin{figure}[!htb]
       \centering
                \includegraphics[width=2.9in]{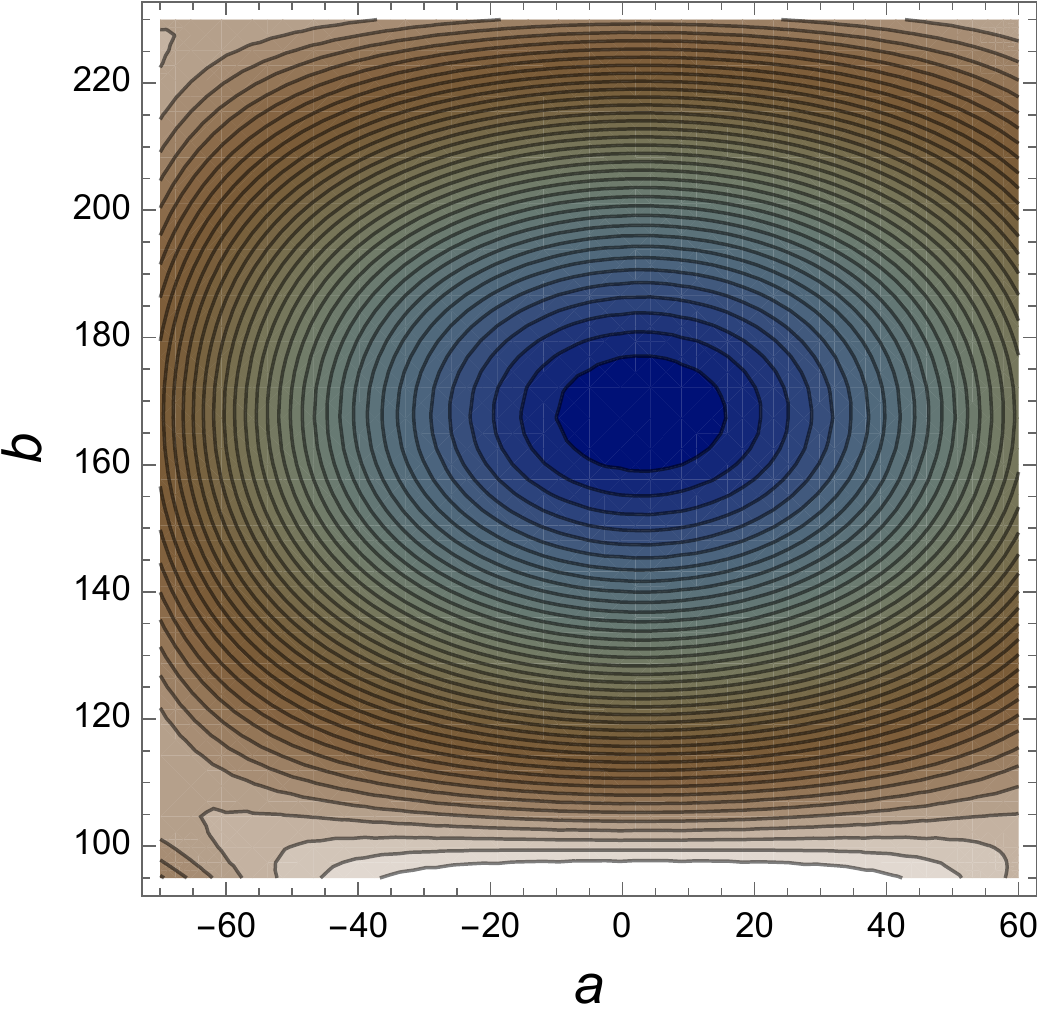}
                \caption{Contour plot of the nuclear configurational entropy (CE) as a function of the free parameters $a$ and $b$. The center of the marine blue ellipsis in the contour plot corresponds to the values of the free parameters $a=2.77$ and $b=166.02$, for the global minimum CE($2.77, 166.02) = 0.476$.}
                \label{fff2}
\end{figure}

Numerically calculated by Eq. (\ref{333}), the nuclear CE  is then plot in Fig. \ref{fff1}. The contour plot in Fig. \ref{fff2} shows that the domain parameter $(a,b)$ can be split into configurational isentropic subdomains, separated by intervals $\Delta {\rm CE}(a,b) = 0.5$. The configurational isentropic subdomains are  
represented by the ellipsoidal-like closed strips, whose boundary are two thin black lines in Fig. \ref{fff2}. The more exterior subdomains correspond to higher values of the nuclear CE, whereas the more inner regions regard lower values of the nuclear CE. The central, marine blue region, surrounds the global minimum of the nuclear CE. 

As it is seen from  Fig. \ref{fff1}, the plot represents the nuclear CE for the baryon production, where the minimum of the plot  coincides with the fitted data from Ref. \cite{Duraes}, with accuracy 1.2\% -- 1.8\%.
The concept which was explored in the present paper, namely, the Shannon's information entropy \cite{Gleiser:2011di} allowed us to find the a global minimum, as a  critical point of the nuclear CE corresponding to the natural choice of the baryon production. This was obtained by the {\tt NMaximize} routine in {\tt Mathematica} of Eq. (\ref{333}), deriving the global minimum CE($2.77, 166.02) = 0.476$, at the point $a=2.77$ and $b=166.02$. These values, solely derived in terms of the nuclear CE, are respectively within 1.8\% of error, for the parameter $a=168$ in Ref. \cite{Duraes}, within 1.2\%,  for the parameter $b=2.82$ also in Ref. \cite{Duraes}.
In such critical point, the interacting system is more stable and dominant, and the calculation of the nuclear CE prove this property.
The results of our numerical calculations show also that outside of the range plot for the observed parameters $a$ and $b$ in Fig. \ref{fff1}, the absence of any other global minima for the baryon production probability.

The computations, summarized in Figs. \ref{fff1} and \ref{fff2}, within the high energy nucleus-nucleus interaction, confirm successfully a global minimum of the CE in the production of baryons at given values of the rapidity.
Such a global minimum of the CE refers to the most dominant state of the nuclear configuration.

One can study other types of nuclear configurations, with other field theoretical effects and other wavefunctions, as the ones proposed in Refs. \cite{daRocha:2005ti,daRocha:2013qhu,Correa:2016pgr,Bazeia:2013usa}. It is our aim to implement also the CE in such a context.

\section{Conclusions}
\label{s4}
At LHC energy the longitudinal momentum fraction carried by a parton in the hadron, the QCD evolution of the CGC describes the RHIC and LHC data for hadron production. The probability of the baryon production is determined in the central and in the forward rapidity regions.
It suggests that in the high energy regime at large rapidities, baryons are produced during the interaction of relatively large fractional momentum partons of the projectile with low fractional momentum partons in the target.
The main changes in the baryon production cross section was interpreted from the point of view the entropy conservation.
As it followed from the CGC formalism, the mechanism of the leading baryon production changes from recombination to independent fragmentation.
The configurational entropy analysis was used in order to justify the leading baryon production in pp/pA/AA collisions at LHC energies.

The results of the study have shown that the global minimum of the nuclear CE, associated to the baryon production probability, in the CGC regime can serve as a
natural choice provided on the comprehensive mathematical and computational analysis. 
Our calculations predicted the values of the free fitted parameters
$a=2.77$ and $b=166.02$ that correspond to the global minimum of the nuclear CE, corroborating and matching HERA data for hadron spectra measured in pp and dAu collisions at RHIC energies, with accuracy between 1.2\% and 1.8\%, respectively for the two free parameters. The errors represent its upper limits for the given fitted parameters for the fixed energy and rapidity values.
We obtain a good agreement with existing data and show predictions for the existing data concerning baryon production mechanism at high energy regime.
The minimum of the nuclear CE  entropy predicts the predominant nuclear states. These phenomena provide the natural onsets of the observables for theoretical and phenomenological predictions as well as for obtained experimental data.

  \acknowledgements
GK thanks to FAPESP (grant No.  2018/19943-6), for partial financial support.

\end{document}